\documentclass[letterpaper,english,aps,prl,twocolumn,notitlepage]{revtex4-2}
\usepackage[T1]{fontenc}
\usepackage[latin9]{inputenc}
\usepackage{amsmath}
\usepackage{amssymb}
\usepackage{graphicx}

\makeatletter

\pdfpageheight\paperheight

\pdfpagewidth\paperwidth

\usepackage{color}
\usepackage[normalem]{ulem}
\usepackage{soul}
\usepackage[usenames,dvipsnames,svgnames,table]{xcolor}
\usepackage{dsfont}
\usepackage{cancel}

\newcommand{\mkadd}[1]{#1}

\newcommand{\eye}{\mathds{1}}

\begin{document}
\title{Optimality of Lindblad unfolding in measurement phase transitions}
\author{Michael Kolodrubetz}
\affiliation{Department of Physics, The University of Texas at Dallas, Richardson,
Texas 75080, USA}
\begin{abstract}
Entanglement phase transitions in hybrid quantum circuits 
describe individual quantum trajectories rather than the
measurement-averaged ensemble, despite the fact that results of measurements
are not conventionally used for feedback. Here, we numerically demonstrate that a class of generalized
measurements with identical measurement-averaged dynamics give different
phases and phase transitions\mkadd{.} 
We show that measurement-averaged destruction
of Bell state entanglement is a useful proxy for determining which hybrid
circuit yields the lowest-entanglement dynamics. We \mkadd{use this to argue that no
unfolding of our model can avoid a volume law phase, which has} implications
for simulation of open quantum systems.
\end{abstract}
\maketitle

Hybrid quantum circuits in which measurements are interspersed with
unitary dynamics have been shown to yield novel non-equilibrium phases
and phase transitions \citep{Li2018,Li2019,Skinner2019,Jian2020,Gullans2020a,Choi2020,Gullans2020,Botzung2021,Buchhold2021,Lavasani2021,Li2021,Li2021a,Alberton2021,Gopalakrishnan2021,Sang2021,Ippoliti2021,Sang2021a,Nahum2021,Turkeshi2021}.
A core concept is that weak or infrequent measurements cut Bell
pairs in a quantum circuit and can decrease entanglement from volume
law to area law. After this was first shown numerically in \citep{Li2018,Skinner2019},
a variety of theoretical perspectives have emerged, including maps
of the circuit dynamics to various statistical mechanics models \citep{Jian2020,Li2021,Li2021a,Ippoliti2021}
and replica tricks in which the steady state maps to the ground state
of an effective Hamiltonian \citep{Bao2020}. Meanwhile, classification
of these phases can be extended to include not just entanglement properties,
but also symmetry breaking, even within the volume law phase \cite{Sang2021}. 

A consistent picture that emerges is that the equilibrium properties
cannot be described by the measurement-averaged density matrix, which
is a featureless infinite temperature state. This is true despite
the fact that the quantities of interest are indeed averaged over
measurements, with no measurement-dependent feedback. It has been
argued that this arises because measurement phases and phase transitions
are only found in quantities that are non-linear in the density
matrix, including Renyi entropies of the (pure state) trajectories.
A complementary perspective is that measurement phases emerge as the
$n\to1$ limit of $n$ replicas, whose measurement-averaged states
encode higher moments of the probability distribution over pure state
density matrices \citep{Bao2020}. 

Despite this perspective, there are nevertheless potential connections
between measurement phase transitions and quantum error correction thresholds
which remain to be understood \citep{Aharonov2000}.
One potential connection comes from the Lindblad equation, which 
is often thought of as a quantum system that is continuously measured by
its environment. Indeed, Lindblad dynamics not only describe
the equilibrium properties of the
steady state, but also its non-equilibrium dynamics through the quantum
regression formula \citep{Lax1963,Lax1967,Carmichael1999}. Such dissipative
dynamics contain information about scrambling 
\cite{Zhang2019, Yoshida2019, GonzalezAlonso2019}, and one
of the perpectives on measurement phase transitions is in terms of
a scrambling and non-scrambling phase \citep{Choi2020}. There remain
many important open cases, such as in what circumstances does measurement-averaged
scrambling dynamics contain information about the underlying measurement
phase transition?

In this paper, we study a family of generalized measurements (``unfoldings'')
such that the measurement-averaged dynamics are identical.
Three conventional unfoldings that we consider give similar entanglement
phase transitions in the steady state, but the exact value of entanglement
and critical measurement strength differs. The fourth unfolding
shows no phase transition, exhibiting a volume law phase independent
of generalized measurement strength. We discuss general properties
for such unfoldings to give different measurement phases and what
general entanglement structure emerges. This result clarifies the
applicability of measurement-averaged dynamics to understand scrambling
and has implications for simulations of open quantum systems, for
which our results imply that different unfoldings of the quantum master
equation lead to different entanglement in the resulting trajectories. While similar results have been seen in the context of free fermion quantum circuits \cite{Cao2019,Piccitto2022}, our results generalize these ideas to the generic non-integrable case, where a volume-law entangled phase is possible.

\begin{figure}
\includegraphics[width=1\columnwidth]{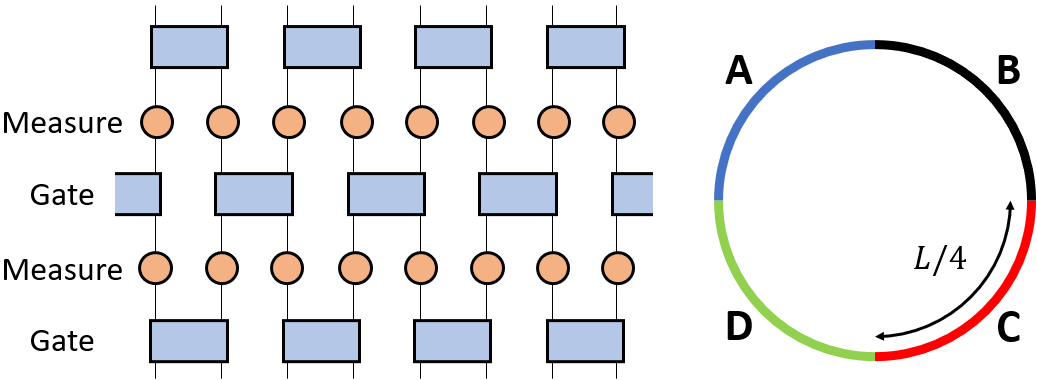}\caption{(left) Illustration of one step of our hybrid circuit model. Boxes
correspond to 2-site random unitaries drawn from the Haar measure.
Circles correspond to generalized $Z$ measurements, defined in the
text. (right) The qubits are arranged on a ring with each quadrant
labeled A-D. }
\label{fig:circuit}
\end{figure}

\emph{Model} -- We consider a quintessential model of measurement
phase transition, in which random 2-qubit Haar unitaries are interspersed
with on-site $Z$-measurements, as illustrated in Figure \ref{fig:circuit}.
For the simplest case of projective measurements with probability
$p$, a number of papers have shown that a phase transition exists
in this model between a volume law entangled phase at low $p$ and
an area law entangled phase at high $p$ \citep{Li2019,Szyniszewski2019,Zabalo2020}.
We can write such measurement dynamics in the language of Kraus operators:
\begin{align*}
M_{0}^{P} & =\sqrt{p}|\uparrow\rangle\langle\uparrow|\\
M_{1}^{P} & =\sqrt{p}|\downarrow\rangle\langle\downarrow|\\
M_{2}^{P} & =\sqrt{1-p}\eye
\end{align*}
For a generic pure state $|\psi\rangle$, each of these outcomes is
obtained with probability $p_{j}=\langle\psi_{j}|\psi_{j}\rangle$,
where $|\psi_{j}\rangle=M_{j}|\psi\rangle$. Averaging over measurement
outcomes, the post-measurement state is given by 
\[
\rho_{f}=\sum_{j}M_{j}\rho_{i}M_{j}^{\dagger}.
\]
For such randomly placed projective measurements, we see that the
result is a pure dephasing channel:
\[
\rho_{f}^{P}=\left(\begin{array}{cc}
\rho_{i,\uparrow\uparrow} & \left(1-p\right)\rho_{i,\uparrow\downarrow}\\
\left(1-p\right)\rho_{i,\downarrow\uparrow} & \rho_{i,\downarrow\downarrow}
\end{array}\right).
\]

The advantage of this Kraus operator formalism is that it can be applied
to generalized measurements. For instance, a simple description of
weak measurements is given by \cite{Li2019}
\begin{align}
M_{0}^{NP} & =\frac{1+\lambda Z}{\sqrt{2\left(1+\lambda^{2}\right)}}\nonumber \\
M_{1}^{NP} & =\frac{1-\lambda Z}{\sqrt{2\left(1+\lambda^{2}\right)}}\label{eq:non_proj_meas}
\end{align}
where the superscript ``NP'' indicates that the measurement is non-projective.
Considering the action on a single qubit, we can again see this corresponds
to a dephasing channel. As shown in the supplement \cite{Supplement_Lindblad_MPT} (which includes reference \cite{Vijay2012}),
\[
\rho_{f}^{NP}=\left(\begin{array}{cc}
\rho_{i,\uparrow\uparrow} & \left(\frac{1-\lambda^{2}}{1+\lambda^{2}}\right)\rho_{i,\uparrow\downarrow}\\
\left(\frac{1-\lambda^{2}}{1+\lambda^{2}}\right)\rho_{i,\downarrow\uparrow} & \rho_{i,\downarrow\downarrow}
\end{array}\right).
\]
Clearly the measurement-averaged dynamics match if $1-p=\frac{1-\lambda^{2}}{1+\lambda^{2}}$,
suggesting that generalized measurement strength $\lambda$ corresponds
to an effective measurement rate
\begin{equation}
p_{eff}^{NP}=\frac{2\lambda^{2}}{1+\lambda^{2}}\label{eq:p_from_lambda}
\end{equation}

As we will see in the next section, both projective and non-projective
measurements behave in a similar way, producing volume law phases
at low $p_{eff}$ and area law at high $p_{eff}$. It might then be
tempting to suggest that the phase transition is indeed \mkadd{identical} for different models of the same measurement-averaged
dynamics. However, we now show that this is not the case by considering
a third generalized measurement protocol, which we refer to as unitary
unfolding. In this case, with probability $q$, the qubit undergoes
a unitary kick with operator $Z$. This is represented by Kraus operators
\begin{align*}
M_{0}^{U} & =\sqrt{q}Z\\
M_{1}^{U} & =\sqrt{1-q}\mathds{1}
\end{align*}
Again, this corresponds to a pure dephasing channel, with identical
measurement-averaged dynamics when 
\begin{equation}
p_{eff}^{U}=2q.
\label{eq:p_eff_U}
\end{equation}
While such unitary kicks do not collect information about the qubit,
they are valid Kraus operators and therefore we refer to this situation
as ``unitary measurements'' and use the superscript ``$U$''
\footnote{Note that, in Eqs.~\ref{eq:p_from_lambda} and \ref{eq:p_eff_U},
one appears to have $p>1$ for $|\lambda| > 1$ or $q > 1/2$. This comes from
the fact that non-projective and unitary protocols cannot be reproduced by
probabilistic projective measurement in this regime. One could reproduce
it by supplementing projective measurement with a deterministic $Z$ gate,
which has no physical effect on entanglement, suggesting that $q$ and $1-q$ 
($\lambda$ and $1/\lambda$) are effectively identical. In this paper, we will
not consider the $p > 1$ regime.}.

Note that the limit of weak continuous measurement corresponds to
Lindblad dynamics, meaning that the strong generalized measurements
above can be generated by finite time evolution under appropriate
unfoldings of the Lindblad equation. Therefore, we refer to these
measurement-averaged dynamics as ``Lindblad equivalent'' and use
the term Lindblad to refer to any such dynamics, even if the measurement
amplitudes are not small.

\begin{figure*}
\includegraphics[width=0.8\textwidth]{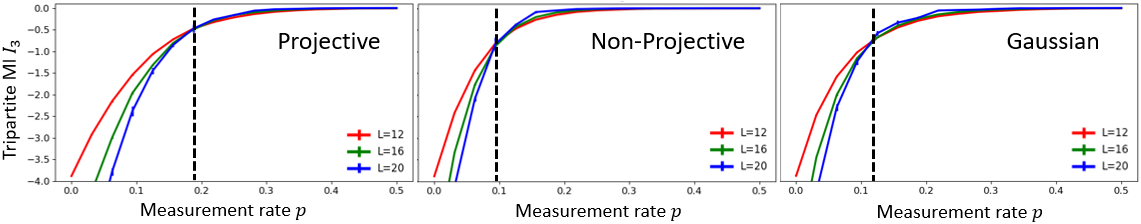}\caption{Comparison of tripartite mutual information (Eq.~\ref{eq:tripartite_mi_definition})
between projective (P), non-projective (NP), and Gaussian (G, see Eq.~\ref{eq:M_G}) measurements
over the same effective range of measurement rate $p$. Dashed lines
show appromimate $p_{c}$ from finite size crossings, which clearly
differs between measurement types.}
\label{fig:entanglement}
\end{figure*}

\emph{Results} -- To confirm these predictions, we numerically examine
the steady state entanglement under these measurement protocols using
exact diagonalization. The conventional measure defining the phase
transition is half-system von Neumann entanglement entropy,
\[
S_{AB}=-\mathrm{Tr}\left[\rho_{AB}\log_{2}\rho_{AB}\right]
\]
where $\rho_{AB}$ is the reduced density matrix of subsystem $AB$,
which has length $L/2$ (see Figure \ref{fig:circuit}). In principle,
$S_{AB}$ is proportional to $L$ in the volume law phase and $O(1)$
in the area law phase. However, entanglement entropy has not been
found to be a sensitive metric for the phase transition. Instead,
we adopt the tripartite mutual information as used in \citep{Zabalo2020}:
\begin{equation}
I_{3}=S_{A}+S_{B}+S_{C}+S_{D}-S_{AB}-S_{BC}-S_{AC}.\label{eq:tripartite_mi_definition}
\end{equation}
While $I_{3}$ is extensive (and negative) in the volume law phase,
it vanishes in the thermodynamic limit within the area law phase,
since boundary contribitions cancel. Therefore it provides much more
useful finite size scaling for detecting the phase transition on small
system size.

The entanglement entropy and tripartite mutual information are seen
in Figure \ref{fig:entanglement}. The unitary unfolding is not shown
explicitily, but for all $p$ it matches the $p=0$ limit of P and
NP measurements.The first thing to note is that neither $S_{AB}$
nor $I_{3}$ matches for the three different unfoldings. This implies
that the steady state ensembles are not microscopically equivalent,
but not necessarily that the phases of matter differ. However, analyzing
crossings of $I_{3}$ clarifies that the three unfoldings indeed give
different phases and phase transitions. Most notably, the unitary
unfolding has \emph{no} phase transition, exhibiting a volume law
phase for arbitrary $q=p/2$. By contrast, both the strong and weak
measurements do exhibit phase transitions. Therefore, we \mkadd{see, as anticipated elsewhere,} that different unfoldings of the measurement-averaged
dynamics \mkadd{generally give different} measurement-induced phase
of matter. \mkadd{We note that} the phase transitions are not guaranteed
to be in the same universality class because, for
instance, the unitary unfolding has no phase transition. \mkadd{Whether 
universality class of the phase transitions can differ away from the unitary
limit is an open question for future work.}

Having established that different unfoldings yield different measurement
phase transitions, it is worth asking the question of which unfolding
works best to minimize entanglement, allowing the area law phase to
survive to the lowest $p_{c}$. To address this, we note that there
is a general trend in the data: non-projective measurement consistently
yields the smallest entanglement entropy, followed by random projective,
and of course unitary measurement has the largest entanglement. This
suggests that, among the measurements considered, non-projective would
be the optimal unfolding for simulation by, e.g., matrix product states. 

\begin{figure}[t]
\begin{centering}
\includegraphics[width=1\columnwidth]{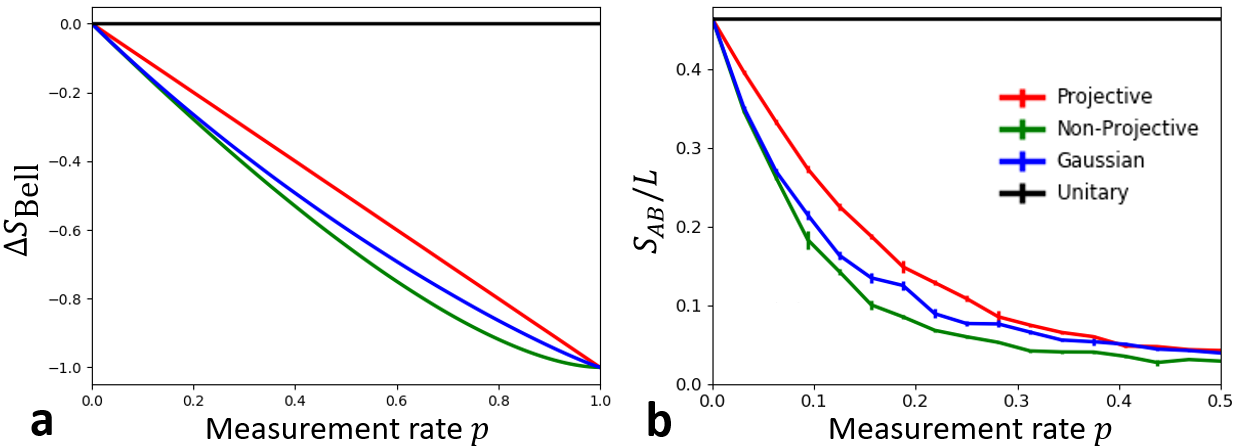}
\par\end{centering}
\caption{(a) Entanglement loss $\Delta S_{\mathrm{Bell}}$ after measurement
of qubit 1 in a Bell pair and (b) steady state half-system entanglement
entropy density $S_{AB}/L$. Many-body entropy $S_{AB}$ lines up
precisely with $\Delta S_{\mathrm{Bell}}$ for all measurements considered,
suggesting $\Delta S_{\mathrm{Bell}}$ as a useful proxy for optimal
unfolding of the measurement-averaged dynamics.}
\label{fig:entanglement_loss}
\end{figure}

Before proceeding to argue that the non-projective measurement specified
in Eq.~\ref{eq:non_proj_meas} is optimal, we need simpler way to
estimate the ability of a given measurement in terms of removing entanglement,
under the assumption that a single measurement that removes entanglement
will result in an overall lower entanglement within the many-body
steady state. We propose a simple test, namely to determine how much
entanglement is lost upon measuring one qubit in a maximally entangled
state, such as the Bell state 
\[
|\psi_{\mathrm{Bell}}\rangle=\frac{|\uparrow\uparrow\rangle+|\downarrow\downarrow\rangle}{\sqrt{2}}.
\]
The loss of entropy of the first qubit $\Delta S_{\mathrm{Bell}}=S_{f}-S_{i}$
is shown for various measurements in Figure \ref{fig:entanglement_loss}.
Clearly it aligns with the results for steady state entropy; a smaller
steady state entropy density corresponds to larger $\left|\Delta S\right|$.
To further test this theory, we consider a slightly more accurate
model of weak measurement in which the histograms of measurement results
are Gaussian distributed with a finite separation between $\uparrow$
and $\downarrow$ corresponding to the measurement strength $\alpha$ \cite{Szyniszewski2019}
\begin{equation}
M^{G}(x)=2^{1/2}\pi^{1/4}\left[G\left(x-\alpha\right)|\uparrow\rangle\langle\uparrow|+G\left(x+\alpha\right)|\downarrow\rangle\langle\downarrow|\right]
\label{eq:M_G}
\end{equation}
where $G(x)$ is a normalized Gaussian of mean $0$ and variance $1$
and $x\in(-\infty,\infty)$ are the possible measurement outcomes.
These Gaussian measurements further support our idea, as both the
Bell state entropy loss $\Delta S_{\mathrm{Bell}}$ and the steady
state entropy $S_{AB}$ are intermediate between non-projective and
projective measurements.

Clearly non-projective measurements outperform random projective measurements
in producing low-entanglement trajectories for the same Lindblad equation,
i.e., are closer to the optimal unfolding for stochastic Schr\"odinger
equation simulations. To argue that the measurements labeled ``NP''
are optimal, we consider the following generic family of generalized
measurements:
\[
M^{gen}(p,x)=\beta_{p}(x)\left[\eye+xZ\right],
\]
a set of non-projective measurements weighted by the real function
$\beta_{p}(x)=\beta_{p}(-x)$. As shown in the supplement \cite{Supplement_Lindblad_MPT}, the function
$\beta_{p}$ is constrained by a normalization condition, $\int_{-\infty}^{\infty}(1+x^{2})\beta_{p}(x)^{2}=1$,
and our goal of matching the measurement-averaged dynamics, which
sets $\int_{-\infty}^{\infty}x^{2}\beta_{p}(x)^{2}dx=p/2$. Note that
all four of the measurement types considered so far fall within this
family with appropriate choices of $\beta_{p}$. In order to better
understand which $\beta_{p}$ will maximize $\left|\Delta S_{\mathrm{Bell}}\right|$,
we start by noticing that, for each $x$, the measurement matches
$M^{NP}(\lambda)$. As seen in Fig.~\ref{fig:entanglement_loss}
and shown analytically in the appendix, Bell state entropy loss is
a convex function in the range $x\in[0,1]$, going from $\Delta S_{\mathrm{Bell}}=0$
at $x=0$ to $\Delta S_{\mathrm{Bell}}=1$ at $x=1$, which corresponds
to a projective measurement. The precise opposite happens for $x>1$,
as $\Delta S(x)=\Delta S(1/x)$. Therefore, the optimal entropy loss
will be given by a $\delta$-function peaked at whatever value is
necessary to match $p$, i.e., the non-projective (NP) measurement.
While this argument is specific to our class of measurements and this
particular system, we expect a similar line of logic to hold in attempting
to determine optimal unfolding of more general Lindblad dynamics.

\emph{Discussion} -- We have shown explicitly that
different unfoldings of the same measurement-averaged (Lindblad-type)
dynamics give rise to different values of entanglement and the entanglement
phase transition in the equivalent hybrid quantum circuit. We find
that destruction of entanglement in a Bell pair is a useful proxy
for many-body steady state entanglement for our class of hybrid circuits.
We use this to show that a non-projective measurement of the form
$\eye\pm\lambda Z$ is optimal for minimizing entanglement.

The most immediate consequences of this work are for numerical simulations
of open quantum systems via the stochastic Schr\"odinger equation,
particularly for entanglement-sensitive methods such as matrix product
states. Our work suggests to use an unfolding of the form $\eye+\lambda Z$
for dephasing channels, which are commonly found experimentally. We
expect that a similar analysis can be applied for other Lindblad operators
as well. Interestingly, our results imply that entanglement
complexity of the stochastic Schr\"odinger equation is not equivalent
to that of the Lindblad evolution, for example by simulating the density
matrix directly as a matrix product operator. In particular, \citep{Noh2020}
showed that for unital quantum channels -- like the ones we examine
here -- density matrices always flow to the area law phase and are
thus efficiently representable. This implies that, for sufficiently
slow Lindblad operators (small $p$), direct Lindblad evolution of
the density matrix is more efficient than stochastic evolution of
even a single pure state trajectory. 
\mkadd{The potential efficiency of density matrix evolution over
trajectories was noted in \citep{Skinner2019}; this work adds to the picture
by arguing that no trajectory unfolding can be as efficient as density matrix evolution.}

In the longer term, this work may provide an interesting perspective on
open quantum systems directly. In particular, the circuit models 
studied here are similar to models of noisy quantum devices, 
with environment playing the role of measurement, for which quantum 
error correction displays phase transitions at finite error rate \citep{Aharonov2000}.
It is clear from our results that no direct connection exists between measurement
phase transitions and error correction in general, as error correction schemes must
handle open quantum systems, e.g., Lindblad dynamics, whose measurement-induced
phases behave differently for different unfoldings. However, there are clear
similarities between these schemes which remain to be explored \mkadd{(cf. \citep{LiArxiv2021}).}
\mkadd{Further discussion of} the general case in which syndrome
measurement combined with environmental dissipation and error-correcting
feedback can be interpreted through the lens of measurement phase transitions
\mkadd{will be the subject of future work}.

\emph{Acknowledgments} -- I am grateful for valuable discussions
with Anushya Chandran, Sarang Gopalakrishnan, Tom
Iadecola, Matteo Ippoliti, Vedika
Khemani, Jed Pixley, Sagar Vijay, Justin Wilson, and Aidan Zahalo.
I am particularly indebted to Aidan for sharing his code and Matteo
for first suggesting the unitary measurement counterexample. This
work was supported by the National Science Foundation through award
number DMR-1945529 and the Welch Foundation through award number AT-2036-20200401.
Part of this work was performed at the Aspen Center for Physics, which
is supported by National Science Foundation grant PHY-1607611, and
at the Kavli Institute for Theoretical Physics, which is supported
by the National Science Foundation under Grant No. NSF PHY-1748958.
Computational resources used include the Frontera cluster operated
by the Texas Advanced Computing Center at the University of Texas
at Austin and the Ganymede cluster operated by the University of Texas
at Dallas' Cyberinfrastructure \& Research Services Department.

\bibliography{References}

\newpage

\onecolumngrid

\section*{Supplementary Information}

In this supplementary information, we derive measurement-averaged
channels and Bell state entropy loss for each of the generalized measurements
considered in the main text. We start by introducing a replica picture
approach which also provides useful notation for doing measurement
averages. We then derive the behavior of each generalized measurement
using this machinery. 

\section{Replica Picture}

As shown in \citep{Bao2020}, the measurement phase transition can
be obtained in the replica limit of a generalized Renyi entropy in
which the vectorized density matrix is averaged over measurement outcomes
in the replica picture. The advantage is that this reduces the trajectory-averaged
entanglement entropy calculation, which is non-linear in density matrix,
to a problem which is linear in replicated density matrix and, thus,
more simple to calculate. As a warmup, consider the superoperator
representation of the unreplicated state. For the case of a projective
$Z$ measurement performed with probability $p$, the final density
matrix can be written as
\[
|\rho_{f}^{(1)}\rangle\rangle=\underbrace{\left[\left(1-p\right)\eye_{1}\eye_{\bar{1}}+p\Pi_{\uparrow1}\Pi_{\uparrow\bar{1}}+p\Pi_{\downarrow1}\Pi_{\downarrow\bar{1}}\right]}_{\mathcal{T}}|\rho_{i}^{(1)}\rangle\rangle
\]
where $\Pi_{\uparrow}=|\uparrow\rangle\langle\uparrow|$ and the subscript
$1$ ($\bar{1}$) indicates acting on the ket (bra). Rearranging,
the transfer matrix $\mathcal{T}$ can be written
\begin{align}
\mathcal{T}_{P}^{(1)} & =\left(1-p\right)\eye_{1}\eye_{\bar{1}}+\frac{p}{4}\left(\eye_{1}+Z_{1}\right)\left(\eye_{\bar{1}}+Z_{\bar{1}}\right)+\frac{p}{4}\left(\eye_{1}-Z_{1}\right)\left(\eye_{\bar{1}}-Z_{\bar{1}}\right)\nonumber \\
 & =\left(1-\frac{p}{2}\right)\eye_{1}\eye_{\bar{1}}+\frac{p}{2}Z_{1}Z_{\bar{1}}\label{eq:transfer_single_replica_projective}
\end{align}
Note that this is precisely the action of a single-qubit dephasing
channel, which we will match for other generalized measurements in
the following sections.

Now we do the same thing for the $n=2$ replica case:
\begin{align*}
\mathcal{T}_{P}^{(2)} & =\left(1-p\right)^{2}\eye_{1}\eye_{\bar{1}}\eye_{2}\eye_{\bar{2}}+\frac{p^{2}}{16}\left(\eye_{1}+Z_{1}\right)\left(\eye_{\bar{1}}+Z_{\bar{1}}\right)\left(\eye_{2}+Z_{2}\right)\left(\eye_{\bar{2}}+Z_{\bar{2}}\right)\\
 & \;\;+\frac{p^{2}}{16}\left(\eye_{1}-Z_{1}\right)\left(\eye_{\bar{1}}-Z_{\bar{1}}\right)\left(\eye_{2}-Z_{2}\right)\left(\eye_{\bar{2}}-Z_{\bar{2}}\right)\\
 & =\left[1-2p+\frac{9p^{2}}{8}\right]\eye_{1}\eye_{\bar{1}}\eye_{2}\eye_{\bar{2}}+\frac{p^{2}}{8}\left(Z_{1}Z_{\bar{1}}+Z_{1}Z_{2}+Z_{1}Z_{\bar{2}}+Z_{\bar{1}}Z_{2}+Z_{\bar{1}}Z_{\bar{2}}+Z_{2}Z_{\bar{2}}\right)\\
 & \;\;+\frac{p^{2}}{8}Z_{1}Z_{\bar{1}}Z_{2}Z_{\bar{2}}\\
\mathcal{T}_{NP}^{(2)} & =\frac{1}{4\left(1+\lambda^{2}\right)^{2}}\left(\eye_{1}+\lambda Z_{1}\right)\left(\eye_{\bar{1}}+\lambda Z_{\bar{1}}\right)\left(\eye_{2}+\lambda Z_{2}\right)\left(\eye_{\bar{2}}+\lambda Z_{\bar{2}}\right)+\\
 & \;\;\frac{1}{4\left(1+\lambda^{2}\right)^{2}}\left(\eye_{1}-\lambda Z_{1}\right)\left(\eye_{\bar{1}}-\lambda Z_{\bar{1}}\right)\left(\eye_{2}-\lambda Z_{2}\right)\left(\eye_{\bar{2}}-\lambda Z_{\bar{2}}\right)\\
 & =\frac{1}{2\left(1+\lambda^{2}\right)^{2}}\Bigg[\eye_{1}\eye_{\bar{1}}\eye_{2}\eye_{\bar{2}}+\lambda^{2}\left(Z_{1}Z_{\bar{1}}+Z_{1}Z_{2}+Z_{1}Z_{\bar{2}}+Z_{\bar{1}}Z_{2}+Z_{\bar{1}}Z_{\bar{2}}+Z_{2}Z_{\bar{2}}\right)\\
 & \;\;+\lambda^{4}Z_{1}Z_{\bar{1}}Z_{2}Z_{\bar{2}}\Bigg]\\
\mathcal{T}_{U}^{(2)} & =\left(1-q\right)^{2}\eye_{1}\eye_{\bar{1}}\eye_{2}\eye_{\bar{2}}+q^{2}Z_{1}Z_{\bar{1}}Z_{2}Z_{\bar{2}}\\
\mathcal{T}_{G}^{(2)} & =\pi\int dx\Bigg\{\left[G\left(x-\lambda\right)\left(\eye_{1}+Z_{1}\right)+G\left(x+\lambda\right)\left(\eye_{1}-Z_{1}\right)\right]\times\\
 & \;\;\;\;\;\;\left[G\left(x-\lambda\right)\left(\eye_{\bar{1}}+Z_{\bar{1}}\right)+G\left(x+\lambda\right)\left(\eye_{\bar{1}}-Z_{\bar{1}}\right)\right]\times\\
 & \;\;\;\;\;\;\left[G\left(x-\lambda\right)\left(\eye_{2}+Z_{2}\right)+G\left(x+\lambda\right)\left(\eye_{2}-Z_{2}\right)\right]\times\\
 & \;\;\;\;\;\;\left[G\left(x-\lambda\right)\left(\eye_{\bar{2}}+Z_{\bar{2}}\right)+G\left(x+\lambda\right)\left(\eye_{\bar{2}}-Z_{\bar{2}}\right)\right]\Bigg\}\\
 & =\pi\int dx\left[G\left(x-\lambda\right)\right]^{4}\eye_{1}\eye_{\bar{1}}\eye_{2}\eye_{\bar{2}}+\\
 & \;\;\pi\int dx\left[G\left(x-\lambda\right)\right]^{2}\left[G\left(x+\lambda\right)\right]^{2}\left(Z_{1}Z_{\bar{1}}+Z_{1}Z_{2}+Z_{1}Z_{\bar{2}}+Z_{\bar{1}}Z_{2}+Z_{\bar{1}}Z_{\bar{2}}+Z_{2}Z_{\bar{2}}\right)+\\
 & \;\;\pi\int dx\left[G\left(x+\lambda\right)\right]^{4}Z_{1}Z_{\bar{1}}Z_{2}Z_{\bar{2}}\\
\int dx\left[G\left(x-\lambda\right)\right]^{4} & =\int dx\frac{G(2\left(x-\lambda\right))}{2\pi}=\frac{1}{4\pi}\\
\int dx\left[G\left(x-\lambda\right)\right]^{2}\left[G\left(x+\lambda\right)\right]^{2} & =\frac{1}{4\pi^{2}}\int dx\exp\left[-\frac{\left(x-\lambda\right)^{2}}{2}-\frac{\left(x+\lambda\right)^{2}}{2}\right]\\
 & =\frac{1}{4\pi^{2}}\int dx\exp\left[-x^{2}-\lambda^{2}\right]=\frac{e^{-\lambda^{2}}}{4\pi^{3/2}}\\
\mathcal{T}_{G}^{(2)} & =\frac{1}{4}\left[\eye_{1}\eye_{\bar{1}}\eye_{2}\eye_{\bar{2}}+\frac{e^{-\lambda^{2}}}{\sqrt{\pi}}\left(Z_{1}Z_{\bar{1}}+Z_{1}Z_{2}+Z_{1}Z_{\bar{2}}+Z_{\bar{1}}Z_{2}+Z_{\bar{1}}Z_{\bar{2}}+Z_{2}Z_{\bar{2}}\right)+Z_{1}Z_{\bar{1}}Z_{2}Z_{\bar{2}}\right]
\end{align*}
An interesting fact, which will hold at higher $n$ as well, is that
the unitary case does not contain the pairwise $Z_{i}Z_{j}$ terms,
while both the projective and non-projective case do. However, coefficients
for projective versus non-projective are different, which will result
in differences between their entanglement. For instance, at leading
order in $p,\lambda\ll1$, the projective case would have both the
pairwise and four ``qubit'' ($Z_{1}Z_{\bar{1}}Z_{2}Z_{\bar{2}}$)
terms, whereas the non-projective case would just include the pairwise
term. Thus, even in this Lindblad limit, the projective and non-projective
cases are different for $n=2$ replicas.

In order to understand the role of such measurements in disentangling
the system within this replica picture, let's consider the general
case
\[
\mathcal{T}^{(2)}\sim\eye_{1}\eye_{\bar{1}}\eye_{2}\eye_{\bar{2}}+\alpha\left(Z_{1}Z_{\bar{1}}+Z_{1}Z_{2}+Z_{1}Z_{\bar{2}}+Z_{\bar{1}}Z_{2}+Z_{\bar{1}}Z_{\bar{2}}+Z_{2}Z_{\bar{2}}\right)+\beta Z_{1}Z_{\bar{1}}Z_{2}Z_{\bar{2}}
\]
acting on a Bell state $|\psi\rangle=\left(|\uparrow\uparrow\rangle+|\downarrow\downarrow\rangle\right)/\sqrt{2}$.
Let's start by introducing notation for our replicated states:
\begin{align*}
|\rho_{i}^{(2)}\rangle\rangle & =\frac{1}{4}\left(|\uparrow\uparrow\rangle+|\downarrow\downarrow\rangle\right)\otimes\left(\langle\uparrow\uparrow|+\langle\downarrow\downarrow|\right)\otimes\left(|\uparrow\uparrow\rangle+|\downarrow\downarrow\rangle\right)\otimes\left(\langle\uparrow\uparrow|+\langle\downarrow\downarrow|\right)\\
 & \equiv\frac{1}{4}\left(|\uparrow\uparrow\uparrow\uparrow\uparrow\uparrow\uparrow\uparrow\rangle\rangle+|\uparrow\uparrow\downarrow\downarrow\uparrow\uparrow\uparrow\uparrow\rangle\rangle+|\uparrow\uparrow\uparrow\uparrow\downarrow\downarrow\uparrow\uparrow\rangle\rangle+\cdots+|\downarrow\downarrow\downarrow\downarrow\downarrow\downarrow\downarrow\downarrow\rangle\rangle\right)
\end{align*}
where we have defined the replicated density matrix supervector via
eigenstates of the basis
\[
|Z_{1,1}Z_{2,1}Z_{1,\bar{1}}Z_{2,\bar{1}}Z_{1,2}Z_{2,2}Z_{1,\bar{2}}Z_{2,\bar{2}}\rangle\rangle
\]
where $Z_{i,j}$ corresponds to replica $j$ of spin at site $i$.
The initial density matrix contains 16 terms in which spin 1 and 2
are aligned independently for each replica and bra/ket. For comparison,
the identity state corresponds to $\eye$ for each spin and replica
independently. In our language, this corresponds to an equal superposition
over spin states in which the bra and the ket ($j$ and $\bar{j}$)
match,
\[
|I^{(2)}\rangle\rangle=|\uparrow\uparrow\uparrow\uparrow\uparrow\uparrow\uparrow\uparrow\rangle\rangle+|\downarrow\uparrow\downarrow\uparrow\uparrow\uparrow\uparrow\uparrow\rangle\rangle+|\uparrow\downarrow\uparrow\downarrow\uparrow\uparrow\uparrow\uparrow\rangle\rangle+\cdots+|\downarrow\downarrow\downarrow\downarrow\downarrow\downarrow\downarrow\downarrow\rangle\rangle
\]
up to an overall prefactor that we define as $1$. Consider the pairwise
and quartic ``interactions'':
\begin{align*}
\mathcal{T}_{pair}^{(2)}|\rho_{i}^{(2)}\rangle\rangle & =\frac{1}{4}\Bigg[6|\uparrow\uparrow\uparrow\uparrow\uparrow\uparrow\uparrow\uparrow\rangle\rangle+0|\uparrow\uparrow\uparrow\uparrow\uparrow\uparrow\downarrow\downarrow\rangle\rangle+0|\uparrow\uparrow\uparrow\uparrow\downarrow\downarrow\uparrow\uparrow\rangle\rangle-2|\uparrow\uparrow\uparrow\uparrow\downarrow\downarrow\downarrow\downarrow\rangle\rangle+\\
 & \;\;\;\;\;\;0|\uparrow\uparrow\downarrow\downarrow\uparrow\uparrow\uparrow\uparrow\rangle\rangle-2|\uparrow\uparrow\downarrow\downarrow\uparrow\uparrow\downarrow\downarrow\rangle\rangle-2|\uparrow\uparrow\downarrow\downarrow\downarrow\downarrow\uparrow\uparrow\rangle\rangle+0|\uparrow\uparrow\downarrow\downarrow\downarrow\downarrow\downarrow\downarrow\rangle\rangle+\\
 & \;\;\;\;\;\;0|\downarrow\downarrow\uparrow\uparrow\uparrow\uparrow\uparrow\uparrow\rangle\rangle-2|\downarrow\downarrow\uparrow\uparrow\uparrow\uparrow\downarrow\downarrow\rangle\rangle-2|\downarrow\downarrow\uparrow\uparrow\downarrow\downarrow\uparrow\uparrow\rangle\rangle+0|\downarrow\downarrow\uparrow\uparrow\downarrow\downarrow\downarrow\downarrow\rangle\rangle-\\
 & \;\;\;\;\;\;2|\downarrow\downarrow\downarrow\downarrow\uparrow\uparrow\uparrow\uparrow\rangle\rangle+0|\downarrow\downarrow\downarrow\downarrow\uparrow\uparrow\downarrow\downarrow\rangle\rangle+0|\downarrow\downarrow\downarrow\downarrow\downarrow\downarrow\uparrow\uparrow\rangle\rangle+6|\downarrow\downarrow\downarrow\downarrow\downarrow\downarrow\downarrow\downarrow\rangle\rangle\Bigg]\\
\mathcal{T}_{quart}^{(2)}|\rho_{i}^{(2)}\rangle\rangle & =\frac{1}{4}\Bigg[|\uparrow\uparrow\uparrow\uparrow\uparrow\uparrow\uparrow\uparrow\rangle\rangle-|\uparrow\uparrow\uparrow\uparrow\uparrow\uparrow\downarrow\downarrow\rangle\rangle-|\uparrow\uparrow\uparrow\uparrow\downarrow\downarrow\uparrow\uparrow\rangle\rangle+|\uparrow\uparrow\uparrow\uparrow\downarrow\downarrow\downarrow\downarrow\rangle\rangle-\\
 & \;\;\;\;\;\;|\uparrow\uparrow\downarrow\downarrow\uparrow\uparrow\uparrow\uparrow\rangle\rangle+|\uparrow\uparrow\downarrow\downarrow\uparrow\uparrow\downarrow\downarrow\rangle\rangle+|\uparrow\uparrow\downarrow\downarrow\downarrow\downarrow\uparrow\uparrow\rangle\rangle-|\uparrow\uparrow\downarrow\downarrow\downarrow\downarrow\downarrow\downarrow\rangle\rangle-\\
 & \;\;\;\;\;\;|\downarrow\downarrow\uparrow\uparrow\uparrow\uparrow\uparrow\uparrow\rangle\rangle+|\downarrow\downarrow\uparrow\uparrow\uparrow\uparrow\downarrow\downarrow\rangle\rangle+|\downarrow\downarrow\uparrow\uparrow\downarrow\downarrow\uparrow\uparrow\rangle\rangle-|\downarrow\downarrow\uparrow\uparrow\downarrow\downarrow\downarrow\downarrow\rangle\rangle+\\
 & \;\;\;\;\;\;|\downarrow\downarrow\downarrow\downarrow\uparrow\uparrow\uparrow\uparrow\rangle\rangle-|\downarrow\downarrow\downarrow\downarrow\uparrow\uparrow\downarrow\downarrow\rangle\rangle-|\downarrow\downarrow\downarrow\downarrow\downarrow\downarrow\uparrow\uparrow\rangle\rangle+|\downarrow\downarrow\downarrow\downarrow\downarrow\downarrow\downarrow\downarrow\rangle\rangle\Bigg]
\end{align*}
Putting these terms together,
\begin{align*}
|\rho_{f}^{(2)}\rangle\rangle & =\frac{1}{4}\Bigg[\left(1+6\alpha+\beta\right)|\uparrow\uparrow\uparrow\uparrow\uparrow\uparrow\uparrow\uparrow\rangle\rangle+\left(1-\beta\right)|\uparrow\uparrow\uparrow\uparrow\uparrow\uparrow\downarrow\downarrow\rangle\rangle+\left(1-\beta\right)|\uparrow\uparrow\uparrow\uparrow\downarrow\downarrow\uparrow\uparrow\rangle\rangle+\left(1-2\alpha+\beta\right)|\uparrow\uparrow\uparrow\uparrow\downarrow\downarrow\downarrow\downarrow\rangle\rangle+\\
 & \;\;\;\;\;\;\left(1-\beta\right)|\uparrow\uparrow\downarrow\downarrow\uparrow\uparrow\uparrow\uparrow\rangle\rangle+\left(1-2\alpha+\beta\right)|\uparrow\uparrow\downarrow\downarrow\uparrow\uparrow\downarrow\downarrow\rangle\rangle+\left(1-2\alpha+\beta\right)|\uparrow\uparrow\downarrow\downarrow\downarrow\downarrow\uparrow\uparrow\rangle\rangle+\left(1-\beta\right)|\uparrow\uparrow\downarrow\downarrow\downarrow\downarrow\downarrow\downarrow\rangle\rangle+\\
 & \;\;\;\;\;\;\left(1-\beta\right)|\downarrow\downarrow\uparrow\uparrow\uparrow\uparrow\uparrow\uparrow\rangle\rangle+\left(1-2\alpha+\beta\right)|\downarrow\downarrow\uparrow\uparrow\uparrow\uparrow\downarrow\downarrow\rangle\rangle+\left(1-2\alpha+\beta\right)|\downarrow\downarrow\uparrow\uparrow\downarrow\downarrow\uparrow\uparrow\rangle\rangle+\left(1-\beta\right)|\downarrow\downarrow\uparrow\uparrow\downarrow\downarrow\downarrow\downarrow\rangle\rangle+\\
 & \;\;\;\;\;\;\left(1+\beta\right)|\downarrow\downarrow\downarrow\downarrow\uparrow\uparrow\uparrow\uparrow\rangle\rangle+\left(1-\beta\right)|\downarrow\downarrow\downarrow\downarrow\uparrow\uparrow\downarrow\downarrow\rangle\rangle+\left(1-\beta\right)|\downarrow\downarrow\downarrow\downarrow\downarrow\downarrow\uparrow\uparrow\rangle\rangle+\left(1+6\alpha+\beta\right)|\downarrow\downarrow\downarrow\downarrow\downarrow\downarrow\downarrow\downarrow\rangle\rangle\Bigg]
\end{align*}
The entanglement is calculated as the replica limit of
\begin{align*}
S_{A}^{(n)} & =\frac{1}{1-n}\log_{2}\left[\frac{\langle\langle I|C_{A}^{(n)}|\rho^{(n)}\rangle\rangle}{\langle\langle I|\rho^{(n)}\rangle\rangle}\right]
\end{align*}
where $C_{A}^{(n)}$ permutes replicas in subsystem $A$. Consider
our case for $A$ given by site 1. 
\begin{align*}
C_{A}^{(2)}|\rho_{i}^{(2)}\rangle\rangle & =\frac{1}{4}\Bigg[|\uparrow\uparrow\uparrow\uparrow\uparrow\uparrow\uparrow\uparrow\rangle\rangle+|\uparrow\uparrow\uparrow\uparrow\uparrow\uparrow\downarrow\downarrow\rangle\rangle+|\downarrow\uparrow\uparrow\uparrow\uparrow\downarrow\uparrow\uparrow\rangle\rangle+|\downarrow\uparrow\uparrow\uparrow\uparrow\downarrow\downarrow\downarrow\rangle\rangle+\\
 & \;\;\;\;\;\;|\uparrow\uparrow\downarrow\downarrow\uparrow\uparrow\uparrow\uparrow\rangle\rangle+|\uparrow\uparrow\downarrow\downarrow\uparrow\uparrow\downarrow\downarrow\rangle\rangle+|\downarrow\uparrow\downarrow\downarrow\uparrow\downarrow\uparrow\uparrow\rangle\rangle+|\downarrow\uparrow\downarrow\downarrow\uparrow\downarrow\downarrow\downarrow\rangle\rangle+\\
 & \;\;\;\;\;\;|\uparrow\downarrow\uparrow\uparrow\downarrow\uparrow\uparrow\uparrow\rangle\rangle+|\uparrow\downarrow\uparrow\uparrow\downarrow\uparrow\downarrow\downarrow\rangle\rangle+|\downarrow\downarrow\uparrow\uparrow\downarrow\downarrow\uparrow\uparrow\rangle\rangle+|\downarrow\downarrow\uparrow\uparrow\downarrow\downarrow\downarrow\downarrow\rangle\rangle+\\
 & \;\;\;\;\;\;|\uparrow\downarrow\downarrow\downarrow\downarrow\uparrow\uparrow\uparrow\rangle\rangle+|\uparrow\downarrow\downarrow\downarrow\downarrow\uparrow\downarrow\downarrow\rangle\rangle+|\downarrow\downarrow\downarrow\downarrow\downarrow\downarrow\uparrow\uparrow\rangle\rangle+|\downarrow\downarrow\downarrow\downarrow\downarrow\downarrow\downarrow\downarrow\rangle\rangle\Bigg]\\
C_{A}^{(2)}|\rho_{f}^{(2)}\rangle\rangle & =\frac{1}{4}\Bigg[\left(1+6\alpha+\beta\right)|\uparrow\uparrow\uparrow\uparrow\uparrow\uparrow\uparrow\uparrow\rangle\rangle+\left(1-\beta\right)|\uparrow\uparrow\uparrow\uparrow\uparrow\uparrow\downarrow\downarrow\rangle\rangle+\left(1-\beta\right)|\downarrow\uparrow\uparrow\uparrow\uparrow\downarrow\uparrow\uparrow\rangle\rangle+\left(1+\beta\right)|\downarrow\uparrow\uparrow\uparrow\uparrow\downarrow\downarrow\downarrow\rangle\rangle+\\
 & \;\;\;\;\;\;\left(1-\beta\right)|\uparrow\uparrow\downarrow\downarrow\uparrow\uparrow\uparrow\uparrow\rangle\rangle+\left(1+\beta\right)|\uparrow\uparrow\downarrow\downarrow\uparrow\uparrow\downarrow\downarrow\rangle\rangle+\left(1+\beta\right)|\downarrow\uparrow\downarrow\downarrow\uparrow\downarrow\uparrow\uparrow\rangle\rangle+\left(1-\beta\right)|\downarrow\uparrow\downarrow\downarrow\uparrow\downarrow\downarrow\downarrow\rangle\rangle+\\
 & \;\;\;\;\;\;\left(1-\beta\right)|\uparrow\downarrow\uparrow\uparrow\downarrow\uparrow\uparrow\uparrow\rangle\rangle+\left(1+\beta\right)|\uparrow\downarrow\uparrow\uparrow\downarrow\uparrow\downarrow\downarrow\rangle\rangle+\left(1+\beta\right)|\downarrow\downarrow\uparrow\uparrow\downarrow\downarrow\uparrow\uparrow\rangle\rangle+\left(1-\beta\right)|\downarrow\downarrow\uparrow\uparrow\downarrow\downarrow\downarrow\downarrow\rangle\rangle+\\
 & \;\;\;\;\;\;\left(1-2\alpha+\beta\right)|\uparrow\downarrow\downarrow\downarrow\downarrow\uparrow\uparrow\uparrow\rangle\rangle+\left(1-\beta\right)|\uparrow\downarrow\downarrow\downarrow\downarrow\uparrow\downarrow\downarrow\rangle\rangle+\left(1-\beta\right)|\downarrow\downarrow\downarrow\downarrow\downarrow\downarrow\uparrow\uparrow\rangle\rangle+\left(1+6\alpha+\beta\right)|\downarrow\downarrow\downarrow\downarrow\downarrow\downarrow\downarrow\downarrow\rangle\rangle\Bigg]\\
2^{-S_{A,i}^{(2)}} & =\frac{\langle\langle I|C_{A}^{(2)}|\rho_{i}^{(2)}\rangle\rangle}{\langle\langle I|\rho_{i}^{(2)}\rangle\rangle}=\frac{1}{2}\\
2^{-S_{A,f}^{(2)}} & =\frac{2\left(1+6\alpha+\beta\right)}{2\left(1+6\alpha+\beta\right)+2\left(1-2\alpha+\beta\right)}=\frac{1+6\alpha+\beta}{2+4\alpha+2\beta}
\end{align*}
For the fully projective case, $\alpha=\beta=1$, this gives the expected
result $e^{-S_{A}^{(2)}}=1$, i.e., zero entanglement. In general,
this is a useful metric to determine how much entanglement is lost,
on average, when a generalized measurement is performed. One advantage
is that it is more straightforward to calculate than von Neumann entropy.
However, since von Neumann entropy is more commonly used to obtain
the measurement phase transition, we solve it below and focus on it
in the main text.

\section{Random Projective Measurement (P)}

We now proceed to calculate details for the various generalized measurements
calculated in the main text. Let's start with random projective measurement.
This can be made into a generalized measurement performed on each
site by introducing a third measurement operator that does nothing.
Specifically, the complete set of measurement operators is
\begin{align*}
M_{0} & =\sqrt{p}|0\rangle\langle0|\equiv\sqrt{p}\Pi_{0}\\
M_{1} & =\sqrt{p}|1\rangle\langle1|\equiv\sqrt{p}\Pi_{1}\\
M_{2} & =\sqrt{1-p}\mathds{1}
\end{align*}
These (Hermitian) measurement operators clearly satisfy the completeness
relation $\sum_{j}M_{j}^{\dagger}M_{j}=\mathds{1}$. Meanwhile, their
action on the density matrix is that of dephasing:
\begin{align*}
\rho_{f} & =\sum_{j}M_{j}\rho_{i}M_{j}^{\dagger}\\
 & =p\rho_{i,00}|0\rangle\langle0|+p\rho_{i,11}|1\rangle\langle1|+\left(1-p\right)\rho_{i}\\
 & =\left(\begin{array}{cc}
\rho_{i,00} & \left(1-p\right)\rho_{i,01}\\
\left(1-p\right)\rho_{i,10} & \rho_{i,11}
\end{array}\right)
\end{align*}
This is precisely the effect of $T_{2}$ dephasing applied for time
$t$ such that $1-p=e^{-t/T_{2}}$. Note that this is a circuit version
of the Lindblad equation. It could be replaced by actual time-dependent
Lindblad dynamics by simply replacing the measurement step by Lindblad
dynamics for finite time with single Lindblad operator $Z$.

It is also worth noting that the same Lindblad-type action can be
obtained directly through the superoperator formalism. As seen in
Eq.~\ref{eq:transfer_single_replica_projective}, 

\[
\mathcal{T}_{P}^{(1)}=\left(1-p\right)\eye_{1}\eye_{\bar{1}}+\frac{p}{4}\left(\eye_{1}+Z_{1}\right)\left(\eye_{\bar{1}}+Z_{\bar{1}}\right)+\frac{p}{4}\left(\eye_{1}-Z_{1}\right)\left(\eye_{\bar{1}}-Z_{\bar{1}}\right)=\left(1-\frac{p}{2}\right)\eye_{1}\eye_{\bar{1}}+\frac{p}{2}Z_{1}Z_{\bar{1}}.
\]
This is the transfer matrix, i.e., measurement-averaged quantum channel,
that we will match throughout this supplement for various measurement
types.

The final entropy of the Bell state is readily calculated by noting
that it is $0$ if a measurement is done and remains $1$ if no measurement
is done. Hence,
\[
S_{f}^{P}=\left(1-p\right)\left(1\right)+\frac{p}{2}\left(0\right)+\frac{p}{2}\left(0\right)=1-p.
\]

\section{Definite Non-Projective Measurement (NP)}

Next, consider the non-projective (NP) measurements:
\[
M_{\pm}=\frac{1\pm\lambda Z}{\sqrt{2\left(1+\lambda^{2}\right)}}
\]
For these measurements, we have
\begin{align}
\mathcal{T}_{NP}^{(1)} & =\frac{1}{2\left(1+\lambda^{2}\right)}\left(\eye_{1}+\lambda Z_{1}\right)\left(\eye_{\bar{1}}+\lambda Z_{\bar{1}}\right)+\frac{1}{2\left(1+\lambda^{2}\right)}\left(\eye_{1}-\lambda Z_{1}\right)\left(\eye_{\bar{1}}-\lambda Z_{\bar{1}}\right)\nonumber \\
 & =\frac{1}{\left(1+\lambda^{2}\right)}\eye_{1}\eye_{\bar{1}}+\frac{\lambda^{2}}{\left(1+\lambda^{2}\right)}Z_{1}Z_{\bar{1}}\label{eq:transfer_single_replica_np}
\end{align}
Note that Eq.~\ref{eq:transfer_single_replica_projective} and \ref{eq:transfer_single_replica_np}
match if $p=2\lambda^{2}/\left(1+\lambda^{2}\right)$, as expected. 

Next let's calculate $\Delta S_{\mathrm{Bell}}$ for NP measurements.
After measurement result $+$ on qubit $1$,
\begin{align*}
|\psi_{+}\rangle & =M_{+1}|\psi_{\mathrm{Bell}}\rangle\\
 & =\frac{\left(1+\lambda\right)|\uparrow\uparrow\rangle+\left(1-\lambda\right)|\downarrow\downarrow\rangle}{2\left(1+\lambda^{2}\right)}\\
\rho_{f1}(x) & =\frac{1}{2\left(1+\lambda^{2}\right)}\left(\begin{array}{cc}
\left(1+\lambda\right)^{2} & 0\\
0 & \left(1-\lambda\right)^{2}
\end{array}\right)\\
S_{f}(\lambda) & =\frac{\left(1+\lambda\right)^{2}}{2\left(1+\lambda^{2}\right)}\log_{2}\left[\frac{2\left(1+\lambda^{2}\right)}{\left(1+\lambda\right)^{2}}\right]+\frac{\left(1-\lambda\right)^{2}}{2\left(1-\lambda^{2}\right)}\log_{2}\left[\frac{2\left(1-\lambda^{2}\right)}{\left(1-\lambda\right)^{2}}\right]
\end{align*}
The same reduced density matrix results for measurement result $-$,
except with the role of $\uparrow$ and $\downarrow$ reversed. Therefore,
entanglement entropy does not depend on measurement outcome. Note
that 
\begin{align*}
\rho_{f1}(\lambda^{-1}) & =\frac{1}{2\left(1+\lambda^{-2}\right)}\left(\begin{array}{cc}
\left(1+\lambda^{-1}\right)^{2} & 0\\
0 & \left(1-\lambda^{-1}\right)^{2}
\end{array}\right)\\
 & =\frac{1}{2\left(1+\lambda^{-2}\right)}\left(\begin{array}{cc}
\lambda^{-2}\left(\lambda+1\right)^{2} & 0\\
0 & \lambda^{-2}\left(\lambda-1\right)^{2}
\end{array}\right)\\
 & =\frac{1}{2\left(1+\lambda^{2}\right)}\left(\begin{array}{cc}
\left(1+\lambda\right)^{2} & 0\\
0 & \left(1-\lambda\right)^{2}
\end{array}\right)=\rho_{f1}(\lambda)
\end{align*}
so $S_{f}(\lambda)=S_{f}(1/\lambda)$.

\section{Unitary Measurement (U)}

Consider the unitary measurement given by Kraus operators
\begin{align*}
M_{0} & =\sqrt{p_{U}}Z\\
M_{1} & =\sqrt{1-p_{U}}\mathds{1}
\end{align*}
Then
\[
\mathcal{T}_{U}^{(1)}=\left(1-q\right)\eye_{1}\eye_{\bar{1}}+qZ_{1}Z_{\bar{1}}
\]
with matching for $q=p/2$. It is clear without calculation that single
site unitaries do not affect entanglement, so $\Delta S_{\mathrm{Bell}}=0$.

\section{Gaussian Weak Measurement (G)}

A Gaussian model of weak measurement is more consistent with experimental
realizations (cf.~\citep{Vijay2012}) and has been considered in
other papers on the measurement phase transition \citep{Szyniszewski2019}.
The Krauss operators are 
\[
M^{G}(x)=2^{1/2}\pi^{1/4}\left[G\left(x-\alpha\right)\Pi_{\uparrow}+G\left(x+\alpha\right)\Pi_{\downarrow}\right]
\]
where $G(z)\propto\exp\left(-z^{2}/2\right)$ is a normalized Gaussian
of mean zero and standard deviation 1. Noting that
\begin{align*}
G^{2}(z) & =\frac{1}{2\pi}e^{-z^{2}}\\
 & =\frac{1}{\left(\sqrt{2\pi}\right)^{2}}e^{-\left(z\sqrt{2}\right)^{2}/2}\\
 & =\frac{G(z\sqrt{2})}{\sqrt{2\pi}}
\end{align*}
and that the measurement operators are again Hermitian, they satisfy
the appropriate completeness relation:
\begin{align*}
\int_{-\infty}^{\infty}dx\left[2^{1/2}\pi^{1/4}G\left(x-\alpha\right)\right]^{2} & =\frac{2\sqrt{\pi}}{\sqrt{2\pi}}\int_{-\infty}^{\infty}dxG\left(\underbrace{\left(x-\alpha\right)\sqrt{2}}_{z^{\prime}}\right)\\
 & =\sqrt{2}\left(\frac{1}{\sqrt{2}}\right)\int_{-\infty}^{\infty}dz^{\prime}G\left(z^{\prime}\right)\\
 & =1\\
\int_{-\infty}^{\infty}dxM^{G}(x)^{\dagger}M^{G}(x) & =\mathds{1}
\end{align*}
The transfer matrix is
\begin{align*}
\mathcal{T}_{G}^{(1)} & =\int_{-\infty}^{\infty}dxM_{1}^{G}(x)M_{\bar{1}}^{G}(x)\\
 & =2\sqrt{\pi}\int_{-\infty}^{\infty}dx\left[G\left(x-\alpha\right)\Pi_{\uparrow1}+G\left(x+\alpha\right)\Pi_{\downarrow1}\right]\left[G\left(x-\alpha\right)\Pi_{\uparrow\bar{1}}+G\left(x+\alpha\right)\Pi_{\downarrow\bar{1}}\right]\\
2\sqrt{\pi}\int_{-\infty}^{\infty}dxG\left(x-\alpha\right)G\left(x+\alpha\right) & =2\pi^{1/2}\left(\frac{1}{2\pi}\right)\int_{-\infty}^{\infty}dx\exp\left[-\frac{1}{2}\left(x-\alpha\right)^{2}-\frac{1}{2}\left(x6\alpha\right)^{2}\right]\\
 & =\frac{1}{\sqrt{\pi}}\int_{-\infty}^{\infty}dx\exp\left[-x^{2}-\alpha^{2}\right]\\
 & =\exp\left[-\alpha^{2}\right]\\
\mathcal{T}_{G}^{(1)} & =\Pi_{\uparrow1}\Pi_{\uparrow\bar{1}}+e^{-\alpha^{2}}\Pi_{\uparrow1}\Pi_{\downarrow\bar{1}}+e^{-\alpha^{2}}\Pi_{\downarrow1}\Pi_{\uparrow\bar{1}}+\Pi_{\downarrow1}\Pi_{\downarrow\bar{1}}\\
 & =\frac{1}{4}\Bigg[\left(\eye_{1}+Z_{1}\right)\left(\eye_{\bar{1}}+Z_{\bar{1}}\right)+e^{-\alpha^{2}}\left(\eye_{1}+Z_{1}\right)\left(\eye_{\bar{1}}-Z_{\bar{1}}\right)+\\
 & \;\;\;\;\;\;e^{-\alpha^{2}}\left(\eye_{1}-Z_{1}\right)\left(\eye_{\bar{1}}+Z_{\bar{1}}\right)+\left(\eye_{1}-Z_{1}\right)\left(\eye_{\bar{1}}-Z_{\bar{1}}\right)\Bigg]\\
 & =\frac{1}{2}\left[\left(1+e^{-\alpha^{2}}\right)\eye_{1}\eye_{\bar{1}}+\left(1-e^{-\alpha^{2}}\right)Z_{1}Z_{\bar{1}}\right]\\
\implies p & =1-e^{-\alpha^{2}}.
\end{align*}

Calculating the final entropy of the Bell state is possible. Note
that, given the continuum of measurement outcomes $x$, the results
for each state must be weighted by the probability of obtaining that
$x$ value. Therefore,
\begin{align*}
|\psi^{G}\left(x\right)\rangle & =\frac{e^{-\left(x-\alpha\right)^{2}/2}|\uparrow\uparrow\rangle+e^{-\left(x+\alpha\right)^{2}/2}|\downarrow\downarrow\rangle}{\sqrt{e^{-\left(x-\alpha\right)^{2}}+e^{-\left(x+\alpha\right)^{2}}}}\\
S^{G}(x) & =\frac{e^{-\left(x-\alpha\right)^{2}}}{e^{-\left(x-\alpha\right)^{2}}+e^{-\left(x+\alpha\right)^{2}}}\log_{2}\left[\frac{e^{-\left(x-\alpha\right)^{2}}}{e^{-\left(x-\alpha\right)^{2}}+e^{-\left(x+\alpha\right)^{2}}}\right]+\frac{e^{-\left(x+\alpha\right)^{2}}}{e^{-\left(x-\alpha\right)^{2}}+e^{-\left(x+\alpha\right)^{2}}}\log_{2}\left[\frac{e^{-\left(x+\alpha\right)^{2}}}{e^{-\left(x-\alpha\right)^{2}}+e^{-\left(x+\alpha\right)^{2}}}\right]\\
p(x) & =\langle\psi_{i}|M^{2}(x)|\psi_{i}\rangle=2\sqrt{\pi}\left[G\left(x-\alpha\right)^{2}\left(\frac{1}{2}\right)+G\left(x+\alpha\right)^{2}\left(\frac{1}{2}\right)\right]\\
 & =\frac{1}{2\sqrt{\pi}}\left[e^{-\left(x-\alpha\right)^{2}}+e^{-\left(x+\alpha\right)^{2}}\right]\\
S_{f}^{G} & =\int_{-\infty}^{\infty}dxp(x)S^{G}(x)=\frac{1}{2\sqrt{\pi}}\int_{-\infty}^{\infty}dx\left(e^{-\left(x-\alpha\right)^{2}}\log_{2}\left[\frac{e^{-\left(x-\alpha\right)^{2}}}{e^{-\left(x-\alpha\right)^{2}}+e^{-\left(x+\alpha\right)^{2}}}\right]+e^{-\left(x+\alpha\right)^{2}}\log_{2}\left[\frac{e^{-\left(x+\alpha\right)^{2}}}{e^{-\left(x-\alpha\right)^{2}}+e^{-\left(x+\alpha\right)^{2}}}\right]\right)
\end{align*}
The final integral can be calculated numerically, obtaining the results
in Figure 3 of the main text.

\section{Universal Generalized Measurements (gen)}

Finally, we consider a family of generalized measurements that encompasses
all previous ones,
\[
M^{gen}(p,x)=\beta_{p}(x)\left[\eye+xZ\right],
\]
with the restriction $\beta_{p}(-x)=\beta_{p}(x)$ on the real-valued
function $\beta_{p}$. Normalization of the Kraus operators gives
\begin{align*}
\eye & =\int_{-\infty}^{\infty}\left[M^{G}(p,x)\right]^{2}dx\\
 & =\eye\int_{-\infty}^{\infty}(1+x^{2})\beta_{p}(x)^{2}dx\\
\implies\int_{-\infty}^{\infty}(1+x^{2})\beta_{p}(x)^{2} & =1
\end{align*}
To match the Lindblad operator with measurement strength $p$, we
need
\begin{align*}
\mathcal{T}_{gen}^{(1)} & =\int_{-\infty}^{\infty}M_{1}^{G}(p,x)M_{\bar{1}}^{G}(p,x)dx\\
 & =\int_{-\infty}^{\infty}\beta_{p}(x)^{2}\left[\eye_{1}+xZ_{1}\right]\left[\eye_{\bar{1}}+xZ_{\bar{1}}\right]dx\\
 & =\int_{-\infty}^{\infty}\beta_{p}(x)^{2}\left[\eye_{1}\eye_{\bar{1}}+x^{2}Z_{1}Z_{\bar{1}}\right]dx\\
\implies\int_{-\infty}^{\infty}x^{2}\beta_{p}(x)^{2}dx & =\frac{p}{2}
\end{align*}

Note that this generalized measurement encompasses all previous ones.
In particular, we have:
\begin{align*}
\beta_{P}(x)^{2} & =\left(1-p\right)\delta\left(x\right)+\frac{p}{2}\left[\delta\left(x-1\right)+\delta\left(x+1\right)\right]\\
\beta_{NP}(x)^{2} & =\frac{1}{2\left(1+\lambda^{2}\right)}\left[\delta\left(x-\lambda\right)+\delta\left(x+\lambda\right)\right]\\
\beta_{U}(x)^{2} & =\left(1-q\right)\delta\left(x\right)+\lim_{y\to\infty}\frac{q\delta\left(x-y\right)}{(1+y^{2})}
\end{align*}
Gaussian measurements require a bit more work because they also have
continuous outcomes whose values need to be rescaled to match those
from $M^{gen}$. Consider the Gaussian measurement outcome $x^{\prime}$.
Then
\begin{align*}
M^{G}(x^{\prime}) & =2^{1/2}\pi^{1/4}\left[G\left(x-\alpha\right)\Pi_{\uparrow}+G\left(x+\alpha\right)\Pi_{\downarrow}\right]=\beta_{G}(x)\left[\eye+xZ\right]\\
\implies\beta_{G}(x) & =\sqrt{\frac{\pi^{1/2}}{2}}\left[G(x^{\prime}-\alpha)+G(x^{\prime}+\alpha)\right]\\
x\beta_{G}(x) & =\sqrt{\frac{\pi^{1/2}}{2}}\left[G(x^{\prime}-\alpha)-G(x^{\prime}+\alpha)\right]\\
x & =\frac{G(x^{\prime}-\alpha)-G(x^{\prime}+\alpha)}{G(x^{\prime}-\alpha)+G(x^{\prime}+\alpha)}
\end{align*}

\end{document}